\documentclass[12pt,preprint]{aastex}

\usepackage{graphicx,times}
\shorttitle{implement a nonlocal convection model for overshoot mixing}
\shortauthors{Zhang}

\begin{document}

\title{A simple scheme to implement a nonlocal turbulent convection model for the convective overshoot mixing }
\author{Q. S. Zhang\altaffilmark{1,2}}
\email{zqs@ynao.ac.cn(QSZ)}

\altaffiltext{1}{Yunnan Observatories, Chinese Academy of Sciences, Kunming 650011, China.}
\altaffiltext{2}{Key Laboratory for the Structure and Evolution of Celestial Objects, Chinese Academy of Sciences, Kunming, 650011, China.}

\begin{abstract}

The classical 'ballistic' overshoot models show some contradictions and are not consistence with numerical simulations and asteroseismic studies.
Asteroseismic studies imply that overshoot is a weak mixing process.
Diffusion model is suitable to deal with it.
The form of diffusion coefficient in a diffusion model is crucial. Because the overshoot mixing is related to the convective heat transport (i.e., entropy mixing), there should be a similarity between them. A recent overshoot mixing model shows consistence between composition mixing and entropy mixing in overshoot region. A prerequisite to apply the model is to know the dissipation rate of turbulent kinetic energy. The dissipation rate can be worked out by solving turbulent convection models (TCMs). But it is difficult to apply TCMs because of some numerical problems and the enormous time cost. In order to find a convenient way, we have used the asymptotical solution and simplified the TCM to be a single linear equation for turbulent kinetic energy. This linear model is easy to be implemented in the calculations of stellar evolution with ignorable extra time cost. We have tested the linear model in stellar evolution, and have found that the linear model can well reproduce the turbulent kinetic energy profile of full TCM, as well as the diffusion coefficient, abundance profile and the stellar evolutionary tracks. We have also studied the effects of different values of the model parameters and have found that the effect due to the modification of temperature gradient in the overshoot region is slight.

\end{abstract}


\section{Introduction}

The convective motion beyond the boundary of the local linear stability is called the convective overshoot.
The mixing caused by the convective overshoot is a major uncertainty of the current stellar evolutionary theory, since it deeply affects the stellar structure but there is still not any solid and easy to be used theory at present. The traditional treatment of overshoot is based on non-local
mixing length theories, e.g., 'ballistic' models \citep{maed75,bre81,z91}, which show an adiabatically stratified and completely mixed overshoot region with a typical length about $0.2 \sim 0.4 H_P$ where $H_P = - dr / dlnP $ is the local pressure scaleheight. Although non-local mixing length models are easy to be implemented in stellar evolution codes and are widely used, they have some contradictions and they do not have enough spatial resolution to accurately describe the overshoot process \citep{ren87}. A property of non-local mixing length models is that there is a jump of $\nabla$ (temperature gradient) from $\nabla_{ad}$ (the adiabatic temperature gradient) to $\nabla_R$ (the radiative temperature gradient) at the boundary of overshoot region. For the sun, the discontinuity of $\nabla$ predicted by non-local mixing length models leads to a characteristic oscillatory component in the frequencies of solar p-modes \citep{go90}. This has been used to estimated the length of the overshoot region below the solar convection zone, and an upper limit has been found as $0.05H_P$ \citep{RV94,ba94,ba94b,mon94,chr95,ba97}. That is too small compared with the prediction of non-local mixing length models. \citet{chr11} have investigated the temperature gradient profile below the base of the solar convection zone and have found that, in order to improve the agreement between models and helioseismic constraints, we actually need a smooth profile of $\nabla$ which are outside the realm of the ¡®non-local mixing length overshoot models.  The helioseismic study may imply that the downward overshoot region below the base of the solar convection zone can not be completely mixed. Because overshoot mixes both entropy and composition \citep{z13}, efficient entropy and composition mixing lead to $dS/dr=0$ and $dX_i/dr=0$ in the overshoot region. Entropy and composition being constants results in $\nabla=\nabla_{ad}$, just like the case in the convection zone with efficient convective heat transport. In a recent asteroseismic study on KIC 10526294 \citep{ma15}, it is also found that assuming a fully mixed overshoot region above the convective core is not the best choice.

For the convective overshoot such a non-local convection phenomenon, besides non-local mixing length models, there are turbulent convection models (TCMs) which are based on statistical equilibrium equations of auto and cross-correlations of velocity and temperature perturbations (e.g., \citet{xio81,xio85,xio97,can97,can98,den06,li07,can11,li12}). \citeauthor{xio81}'s \citeyearpar{xio81} TCM and \citeauthor{li07}'s \citeyearpar{li07} TCM have been applied in the solar structure models and have been found to provide the required smooth $\nabla$ profile \citep{zl12a,zdxc12}. The temperature gradient profile outside the Schwarzchild local convective boundary predicted by \citeauthor{xio81}'s \citeyearpar{xio81} TCM or \citeauthor{li07}'s \citeyearpar{li07} TCM is different from the prediction of non-local mixing length models. \citet{z91} has proposed to use two word 'penetration' and 'overshoot' to distinguish the convective motions beyond the Schwarzchild boundary with high efficiency of convective heat transport ('penetration') and with low efficiency ('overshoot'), respectively, and the efficiency of penetration convection is so high that the dominated region is nearly adiabatic stratified. This adiabatic penetration convection region has been predicted by the non-local mixing length models, e.g., \citet{maed75,bre81,z91}. In TCMs (e.g., \citet{xio81,li07}), however, that is not the case. TCMs show a continuous profile of temperature gradient with $\nabla_R< \nabla< \nabla_{ad}$ and there is no significant adiabatic 'penetration'. Another property of the TCMs is that $\nabla-\nabla_R$ is significant nonzero only in a thin layer near the Schwarzchild boundary and $\nabla-\nabla_R\approx0$ in the further region in which turbulent kinetic energy is still nonzero. This means that the TCMs predict a thin 'penetration' layer (the convective heat flux is significant nonzero but is not enough to result in $\nabla\approx \nabla_{ad}$) and an thick 'overshoot' region (the convective heat flux is too small to modify $\nabla$). Theoretical analysis shows that the depth of the thin 'penetration' layer predicted by TCMs is about $\sim1H_k$ ($H_k=\mid dr/dlnk \mid$ is the scale height of turbulent kinetic energy) regardless of what values of parameters are adopted \citep{zl12b}. The thick 'overshoot' region can cover many magnitude of order of turbulent kinetic energy. The main reason resulting in such a distinction between TCMs and non-local mixing length models is that the latter include the assumption that vertical velocity and temperature fluctuation is strongly correlated, which results in significant convective heat flux making the stratification outside the convective boundary quasi-adiabatic \citep{pm95}. Numerical solutions of TCMs (e.g., \citet{xio85,xio01,zl12b}) show that the correlation coefficient between vertical velocity and temperature fluctuation changes from 1 to 0 near the convective boundary. A weak correlation between vertical velocity and temperature fluctuation leads to an exponential solution of turbulent kinetic energy and a deep convective overshoot \citep{pm95}, which has been found in numerical solutions of TCMs. Flows penetrate from the convective unstable zone to the convective stable zone (corresponding to \emph{'penetration'} of \emph{'fluid particles'}) could preserve strong correlation and contribute to convective heat flux, but the flows originally located in the stable zone (corresponding to \emph{'overshoot'} of \emph{kinetic energy}) are of weak correlation thus have little contribution to convective heat flux. Those two kinds of flows are the reason of the decreasing correlation between vertical velocity and temperature fluctuation \citep{pm95,z13}. It should be noticed that the latter which is caused by the convective transport of the kinetic energy was usually ignored in non-local mixing length models. The weak correlation between vertical velocity and temperature fluctuation has been confirmed in numerical simulations (e.g., \citet{sin95,ma07}). The numerical simulations of convective penetration and overshoot \citep{bru02} have found that the convective penetration can not establish an adiabatic stratification even though the P\'{e}clet number is much larger than unit. Those results of simulations are inconsistent with non-local mixing length models.

Beside the traditional treatment, we could model the overshoot mixing as a diffusion process (e.g., \citet{den96,ven98,he2000,zl12a,z13}). The point in a diffusion model is the form of the adopted diffusion coefficient. For example, in \citet{ven98}, the diffusion coefficient is $D = u l_d / 3$ where $u$ is characteristic turbulent speed and $l_d$ is the convective scale length, as similar as in the convection zone, and the characteristic turbulent speed $u$ is assumed as exponential decreasing based on \citeauthor{xio85}'s \citeyearpar{xio85} turbulent convection model. This diffusion model predicts an exponential decreasing diffusion coefficient $D$ in overshoot region and the characteristic length for mixing in overshoot region is the same as the characteristic length in convection zone. On the other hand, the form of the diffusion coefficient in overshoot mixing should be related to the convective heat transport in the overshoot region because the latter is actually caused by the entropy mixing. In high P\'{e}clet number overshoot region, the convective time scale is too short for flows to exchange their entropy, as well as their composition. The similarity between composition mixing and entropy mixing implies that the form of both may be the same. Turbulent convection models show that the convective heat flux $\overline {{u_r}'T'}$ in the overshoot region is \citep{xio89,den96,li07,zl12b,li12}:
\begin{eqnarray} \label{utallTCM}
\overline {{u_r}'T'}  \propto  - \frac{T}{{\delta g}}\varepsilon
\end{eqnarray}%
where $g$ is the gravitational acceleration, $T$ is temperature, $\delta= - ({\partial \ln \rho } /{\partial \ln T})_P$ is the dimensionless expansion coefficient and $\varepsilon$ is the dissipation rate of turbulent kinetic energy. Therefore the entropy flux $\overline {{u_r}'S'}$ of the overshoot region is:
\begin{eqnarray} \label{uS}
\overline {{u_r}'S'}  \approx \frac{{{c_P}}}{T}\overline {{u_r}'T'}  \propto  - \frac{{{c_P}}}{{\delta g}}\varepsilon  \sim  - \frac{\varepsilon }{{{N^2}}}\frac{{\partial S}}{{\partial r}}
\end{eqnarray}%
where $N^2$ describes the squared buoyancy frequency, $S$ is entropy and $c_P$ is specific heat capacity at constant pressure. This expression shows that the diffusion coefficient for entropy mixing in high P\'{e}clet number overshoot region is $D_S \propto \varepsilon / N^2$. In \citeauthor{z13}'s \citeyearpar{z13} convective mixing model, the diffusion coefficient is solved based on hydrodynamic equations and some closure assumptions. The solution is that the diffusion coefficient for convective mixing in the convection zone is of the form $D \propto k^2/\varepsilon \sim u l$ and the diffusion coefficient for convective overshoot mixing is of the form $D \propto \varepsilon / N^2$. The result is consistent with the convective entropy mixing in both convection zone and overshoot region.

The prerequisite of applying \citeauthor{z13}'s \citeyearpar{z13} convective mixing model is to know the dissipation rate of turbulent kinetic energy $\varepsilon$ in the overshoot region. At present, an practicable option is to use TCMs (e.g., \citet{xio81,xio85,xio97,can97,can98,den06,li07,can11,li12}) which has been suggested to deal with the convective overshoot by the helioseismic study \citep{chr11}. Those TCMs are based on hydrodynamic equations and closure assumptions, describe the evolution and distribution of averaged correlations of turbulent variables ($\overline {{u_r}'{u_r}'}$, $\overline {{u_r}'T'}$, $\overline {T'T'}$, $\varepsilon$, etc. ) in stellar interior. However, TCMs are highly nonlinear equations, too complicated to be applied in stellar evolution. Sometimes it is difficult to find a solution satisfying both TCM equations and the stellar structure equations due to numerical calculation problems. Even for the converged stellar evolution models, the time cost is enormous (normal time cost multiplying by a factor of $50\sim100$) \citep{z15}. In order to apply the convective mixing model, it is necessary to simplify TCMs to stably and quickly solve the distribution of turbulent kinetic energy $\varepsilon$ in stellar interior. In this paper, we introduce a simple scheme to implement \citeauthor{li07}'s \citeyearpar{li07} TCM for the convective overshoot mixing. The content of this paper is as follows: the overshoot mixing model is introduced in section 2, the TCM and its properties are introduced in section 3, the details of the simple scheme are described in section 4, the numerical results of the simple scheme are shown in section 5 and section 6 is a summary.

\section{The overshoot mixing model}

In this paper, \citeauthor{z13}'s \citeyearpar{z13} model of overshoot mixing is adopted. The model is derived from fluid dynamic equations and some assumptions. The model shows that the convective overshoot mixing in high P\'{e}clet number region can be treated as a diffusion process with the diffusion coefficient as follow:
\begin{eqnarray} \label{z13ovm}
D = {C_{OV}}\frac{\varepsilon }{{{N_{turb}}^2}}
\end{eqnarray}%
where $N_{turb}^2$ is calculated as
\begin{eqnarray} \label{z13ovmn2}
{N_{turb}}^2 =  - \frac{{\delta g}}{{{H_P}}}[\nabla  - {\nabla _{ad}} - \\ \nonumber
{C_1}{C_A}\sum\limits_{i = 1}^I {{{(\frac{{\partial \ln T}}{{\partial {X_i}}})}_{P,\rho ,X - \{ {X_i}\} }}\frac{{d{X_i}}}{{d\ln P}}} ]
\end{eqnarray}%
where $I$ is the number of independent elements, $\varepsilon$ is the dissipation rate of turbulent kinetic energy, $\nabla_{ad}$ is adiabatic temperature gradient, $\nabla$ is real temperature gradient in stellar interior, $C_{OV}$, $C_1$ and $C_A$ are model parameters, $X_i$ is the mass fraction of the i-th element, other symbols are with their usual meanings. The parameter $C_{OV}$ is a proportion factor which could be determined by calibrations of fitting observation, the parameter $C_1$ is used to model the turbulent abundance-abundance correlation $\overline{X_i'X_j'}$ and the parameter $C_A$ is used to model the dissipation of turbulent temperature-abundance correlation $\overline{T'X_j'}$. $N_{turb}^2$ is similar to the squared Brunt-V\"{a}is\"{a}l\"{a} frequency $N^2$ since $N_{turb}^2=N^2$ when $C_1 C_A=1$ which is assumed in \citet{z13}. However, according to \citet{can98} and \citet{can11}, $C_1=\sigma_t=0.72$ where $\sigma_t$ is the turbulent Prandtl number.

The representation of the diffusion coefficient shows the image that the length scale (in radial) for mixing $l_{mix}=\sqrt{k}/N_{turb}$ and the lifetime $\tau_{mix}=\tau=k/\varepsilon$ where $k$ is the turbulent kinetic energy, since the diffusion coefficient is $D \propto l_{mix}^2/\tau_{mix}$.
The diffusion coefficient of matter mixing has the same form to convective heat transport (i.e., entropy mixing) in high P\'{e}clet number overshoot region \citep{z13}, because the turbulent convection models (e.g., \citet{xio89,den06,zl12b}) show that the convective heat flux is proportion to the dissipation rate of turbulent kinetic energy in high P\'{e}clet number overshoot region. The physical reason is that the convective heat transport in high P\'{e}clet number region is equivalent to the entropy mixing and the entropy mixing is an accessory of the matter mixing \citep{z13}. The same form for matter mixing and for convective heat transport implies the consistence between turbulent convection models and the overshoot mixing model.

In order to apply the overshoot mixing model in stellar evolution, one must know the dissipation rate of turbulent kinetic energy, i.e., $\varepsilon$, in the overshoot region. At present, we can calculate the dissipation rate of turbulent kinetic energy in the overshoot region by using turbulent convection models.

\section{ Li \& Yang's (2007) nonlocal turbulent convection model and its properties}

The turbulent convection model (TCM) adopted in this paper was developed by \citet{li07}:
\begin{eqnarray} \label{TCM_kr}
\frac{\partial }{{\partial m}}[{(\frac{{dm}}{{dr}})^2}(2{C_s}{k_r}\tau )\frac{{\partial {k_r}}}{{\partial m}}] = \frac{1}{3}k{\tau ^{ - 1}} - \frac{{\delta g}}{T}\overline {{u_r}'T'} \\ \nonumber
 + {C_k}{\tau ^{ - 1}}({k_r} - \frac{k}{3}),
\end{eqnarray}%
\begin{eqnarray} \label{TCM_k}
\frac{\partial }{{\partial m}}[{(\frac{{dm}}{{dr}})^2}(2{C_s}{k_r}\tau )\frac{{\partial k}}{{\partial m}}] = k{\tau ^{ - 1}} - \frac{{\delta g}}{T}\overline {{u_r}'T'},
\end{eqnarray}%
\begin{eqnarray} \label{TCM_uT}
\frac{\partial }{{\partial m}}[{(\frac{{dm}}{{dr}})^2}(4{C_{t1}}{k_r}\tau )\frac{{\partial \overline {{u_r}'T'} }}{{\partial m}}] =  - \frac{{\delta g}}{T}\overline {T'T'} \\ \nonumber
 - 2{k_r}\frac{T}{{{H_P}}}(\nabla  - {\nabla _{ad}})
  + {C_t}(1 + {P_e}^{ - 1}){\tau ^{ - 1}}\overline {{u_r}'T'},
\end{eqnarray}%
\begin{eqnarray} \label{TCM_TT}
\frac{\partial }{{\partial m}}[{(\frac{{dm}}{{dr}})^2}({C_{e1}}{k_r}\tau )\frac{{\partial \overline {T'T'} }}{{\partial m}}] =  - \overline {{u_r}'T'} \frac{T}{{{H_P}}}(\nabla  - {\nabla _{ad}}) \\ \nonumber
 + {C_e}(1 + {P_e}^{ - 1}){\tau ^{ - 1}}\overline {T'T'}.
\end{eqnarray}%

In above equations, the meanings of symbols are as follows: $k_r=\overline{u_r'u_r'}/2$ is the radial kinetic energy, $\overline{u_r'T'}$ is the convective heat flux, $\overline{T'T'}$ is the temperature variance, $\tau=k/\varepsilon$ is the turbulent dissipation time scale with the turbulent dissipation $\varepsilon=k^{3/2}/l$ in which $l=\alpha H_P$, $P_e=l\sqrt{k}/D_R$ is the Pecl\'{e}t number in which the radiative diffusion coefficient $D_R=\lambda/\rho c_P$ and the thermal conduction coefficient $\lambda=4acT^3/(3\kappa\rho)$, $C_s$, $C_t$, $C_e$, $C_{t1}$, $C_{e1}$, $C_k$ and $\alpha$ are model parameters, other symbols are with their usual meanings. The parameter $C_e$ in this model is related to the overshoot mixing model by $C_{OV}=C_A-C_e$ \citep{z13}.

This TCM has been investigated in theoretical by \citet{zl12b}. Now we recall the main results.

In the convection zone with high P\'{e}clet number, turbulence is nearly in local equilibrium, thus the localized model (ignoring the diffusion terms on the l.h.s. of the equations of the nonlocal model) is reasonable to describe the turbulent convection \citep{li01}. The approximate solution of the localized model in high P\'{e}clet convection zone shows that the temperature gradient is very close to the adiabatic temperature gradient .

In the overshoot region, the diffusion of turbulent kinetic energy is necessary since the turbulent energy in the convective overshoot region is supported by nonlocal convective transport. By ignoring the diffusions of $\overline{u_r'T'}$ and $\overline{T'T'}$ (i.e., setting $C_{t1}=C_{e1}=0$), it has been found that the asymptotical solution in overshoot region with $P_e \gg 1$:
\begin{eqnarray} \label{kOV}
k = {k_C}{(\frac{P}{{{P_C}}})^\theta },
\end{eqnarray}%
\begin{eqnarray} \label{u't'OV}
\overline {{u_r}'T'}  = {\rm{Max}}\{  - \frac{T}{{{H_P}}} D_R ({\nabla _{ad}} - {\nabla _R}), - 2{C_e}\omega \frac{T}{{\delta g}}\varepsilon\},
\end{eqnarray}%
\begin{eqnarray} \label{t't'OV}
\overline {T'T'}  = 2\frac{T}{{{H_P}}}\frac{T}{{\delta g}}({\nabla _{ad}} - \nabla )\omega k,
\end{eqnarray}%
where $k_C$ is $k$ at the Schwarzchild convective boundary where $\nabla_R=\nabla_{ad}$, $\omega=k_r/k$ is the anisotropic degree, $\theta=dlnk/dlnP$ is the exponential decreasing index of turbulent kinetic energy in overshoot region.

The exponential decreasing index of turbulent kinetic energy in overshoot region $\theta$ is determined by:
\begin{eqnarray} \label{ksita}
\theta  =  \pm \frac{1}{{{\alpha}}}\sqrt {\frac{{1 + 2{C_e}{\omega _O}}}{{3{C_s}{\omega _O}}}},
\end{eqnarray}%
where the sign depends on the direction of overshoot: positive for upward and negative for downward.

The value of $k_C$ can be estimated by using the 'the maximum of diffusion' method as:
\begin{eqnarray} \label{kcb}
{k_C}^{\frac{3}{2}} \approx \frac{1}{e}{k_{B,Local}}^{\frac{3}{2}}
\approx \frac{1}{e}{\alpha}{[\delta g{D_R}({\nabla _R} - {\nabla _{ad}})]_B}.
\end{eqnarray}%
where location B is a point in the convection zone with the distance to the convective boundary being:
\begin{eqnarray} \label{rb}
\left| {{r_B} - {r_C}} \right| = \sqrt {\frac{{4{C_s}{\omega _C}}}{3}} l
\end{eqnarray}%
where $r_C$ is the radius at the convective boundary.

Some typical values of the anisotropic degree $\omega$ in some cases are as follows: $\omega_{CZ}$ the anisotropic degree in the convection zone:
\begin{eqnarray} \label{omicz}
{\omega _{CZ}} = \frac{2}{{3{C_k}}} + \frac{1}{3},
\end{eqnarray}%
$\omega_{C}$ the anisotropic degree at the convective boundary:
\begin{eqnarray} \label{omic}
{\omega _C} \approx \frac{1}{2}({\omega _{CZ}} + \frac{1}{3}),
\end{eqnarray}%
and $\omega_{O}$ the asymptotical equilibrium value of the anisotropic degree in overshoot region satisfies the following equation:
\begin{eqnarray} \label{omio}
2{C_e}{\omega _O}^2 - ({C_k} - 1 + 2{C_e}){\omega _O} + \frac{1}{3}({C_k} - 1) = 0.
\end{eqnarray}%
The parameter $C_k$ should be larger than 1 and the smaller root of $\omega _O$ is the physical root (see Appendix A).

The above solution is for the turbulent convection model in overshoot region with $P_e \gg 1$, and the the diffusions of $\overline{u_r'T'}$ and $\overline{T'T'}$ are ignored. In the low $P_e$ overshoot region, it is mathematically required that turbulent variables must be cut off in a short distance. Therefore it is reasonable to ignore the overshoot in low $P_e$ region. The diffusions of $\overline{u_r'T'}$ is ignorable since the diffusion is much less than the local terms. The diffusions of $\overline{T'T'}$ smoothes the profile of $\overline{u_r'T'}$ and $\overline{T'T'}$ near the convective boundary but basically does not affect the profile of $k$. For those reasons, we can use the above approximate / asymptotical solution instead of the numerical solution of the TCM.

\section{Linear model of nonlocal turbulent convection model for overshoot mixing}

By solving \citeauthor{li07}'s \citeyearpar{li07} nonlocal turbulent convection model to obtain the dissipation rate of turbulent kinetic energy $\varepsilon=k^{3/2}/l$ where $l=\alpha H_P$, one can apply the overshoot mixing model. However, it is difficult to apply such a nonlocal turbulent convection model in stellar evolution (e.g., \citet{z15}): the time cost is enormous and numerical instability can not be totally resolved yet.
It is necessary to find a simple approach in order to work out the dissipation rate conveniently. The asymptotical solution is simple to be used but the estimate of $k_C$ can not be used for thin convection zone or a small convective core, and the accuracy of the estimate is not high enough. We need to find a better way.

The equation of turbulent kinetic energy in diffusion equilibrium Eq.(\ref{TCM_k}) is equivalent to the following equation:
\begin{eqnarray} \label{est_kc}
\frac{\partial }{{\partial m}}[{(\frac{{dm}}{{dr}})^2}(\frac{4}{3}{\omega}{C_s}l)\frac{{\partial {k^{\frac{3}{2}}}}}{{\partial m}}] = \varepsilon - \frac{{\delta g}}{T}\overline {{u_r}'T'}.
\end{eqnarray}%
As mentioned above, the temperature gradient in high P\'{e}clet convection zone is near adiabatic temperature gradient, and the convective heat flux in high P\'{e}clet overshoot region satisfies Eq.(\ref{u't'OV}). Therefore the convective heat flux in high P\'{e}clet region (no matter in convection zone or in overshoot region) satisfies Eq.(\ref{u't'OV}):
\begin{eqnarray} \label{u't'linear}
\overline {{u_r}'T'}  = {\rm{Max}}\{  - \frac{T}{{{H_P}}} D_R ({\nabla _{ad}} - {\nabla _R}), - 2{C_e}\omega \frac{T}{{\delta g}}\varepsilon\}.
\end{eqnarray}%
The point of junction at which $- (T/H_P) D_R ({\nabla _{ad}} - {\nabla _R}) = - 2{C_e}\omega (T/\delta g)\varepsilon$ locates in the overshoot region with the distance to the convective boundary being \citep{zl12b}:
\begin{eqnarray} \label{lad}
{l_{ad}} \approx \varphi H_P, \varphi = \frac{{\alpha \sqrt {\frac{{4{C_s}{\omega _C}}}{3}} }}{{\frac{e}{{2{C_e}{\omega _C}}} + 1}}.
\end{eqnarray}%
Therefore the convective heat flux in high P\'{e}clet region can also be written as:
for the case of ${\nabla _{ad}} > {\nabla _R}$ and $\left| {\ln \frac{P}{{{P_C}}}} \right| > {\varphi}$:
\begin{eqnarray} \label{u't'linear-a}
\overline {{u_r}'T'}  =  - 2{C_e}\omega \frac{T}{{\delta g}}\varepsilon,
\end{eqnarray}
where $P_C$ is the pressure of the closest convective boundary, and for other cases:
\begin{eqnarray} \label{u't'linear-b}
\overline {{u_r}'T'} =  - \frac{T}{{{H_P}}} D_R({\nabla _{ad}} - {\nabla _R}).
\end{eqnarray}
Taking the representation of the convective heat flux Eq.(\ref{u't'linear-a}) and Eq.(\ref{u't'linear-b}) into Eq.(\ref{est_kc}), and noting that $\varepsilon=k^{3/2}/l$, we get a linear equation of $k^{3/2}$.

Another variable needed to be determined is the anisotropic degree $\omega$. In the convection zone, $\omega$ changes from $\omega_{C}$ to $\omega_{CZ}$ in the region near the convective boundary with the diffusion of $k$ dominating. In overshoot region, numerical calculations shows that $\omega$ changes from $\omega_{C}$ to $\omega_{O}$ near the convective boundary in a typical length about $1H_k$ where $H_k=|dr/dlnk|=H_P/|\theta|$ is the scale height of turbulent kinetic energy. Thus, we estimate $\omega$ by using linear interpolation as follows:
for the case of ${\nabla _R} \ge {\nabla _{ad}}$:
\begin{eqnarray} \label{omi-a}
\omega  =  {\rm{Min}}(1,\chi ){\omega _{CZ}} + {\rm{Max}}(0,1 - \chi ){\omega _C}, \\ \nonumber
\chi  = \frac{1}{\alpha }\sqrt {\frac{3}{{4{C_s}{\omega _C}}}} \left| {\ln \frac{P}{{{P_C}}}} \right|,
\end{eqnarray}%
and for the case of ${\nabla _R} < {\nabla _{ad}}$:
\begin{eqnarray} \label{omi-b}
\omega  = {\rm{Min}}(1,\beta ){\omega _O} + {\rm{Max}}(0,1 - \beta ){\omega _C}, \\ \nonumber
\beta  = \left| {\frac{1}{\theta }\ln \frac{P}{{{P_C}}}} \right|,
\end{eqnarray}%
where $P_C$ is the pressure of the closest convective boundary.

A linear model of turbulent kinetic energy in diffusion equilibrium with $P_e \gg 1$ comprises Eqs. (\ref{est_kc}), (\ref{u't'linear-a}), (\ref{u't'linear-b}), (\ref{omi-a}) and (\ref{omi-b}). In order to solve the linear model, we need to set two boundary conditions. A reasonable set of boundary conditions are zero flux at the stellar center and the stellar surface:
\begin{eqnarray} \label{bc}
{\left. {\frac{{\partial k}}{{\partial m}}} \right|_{m = 0}} = {\left. {\frac{{\partial k}}{{\partial m}}} \right|_{m = M}} = 0.
\end{eqnarray}%
The problem is that, in general, P\'{e}clet number is low in a thin envelope of a star. This leads to some mistakes when the linear model with the assumption $P_e \gg 1$ is used. In thin convective envelope(s) below the stellar surface, the ignorable radiative heat exchange leads to low P\'{e}clet number so that the temperature gradient is higher than the adiabatic temperature gradient. In this case the convective heat flux is smaller than the adiabatic heat flux (e.g., Eq.(\ref{u't'linear-b})) so that the kinetic energy should be smaller than the value determined by the linear model. However, the difference should exist only in the low P\'{e}clet layer extending by several typical diffusion length scale $\sim l$. Therefore using those boundary conditions should not lead to mistake for the convective core overshooting and the thick convective envelope downward overshooting. If the final solution of turbulent kinetic energy shows some regions with $P_e= l \sqrt{k}/D_R < 1$ in overshoot region, we suggest to reset zero turbulent kinetic energy in those region, for the reason that the turbulent convection model shows that the turbulent variables quickly cut-off in low P\'{e}clet region as mentioned in Section 3.

\section{Numerical results}

We use the stellar evolution code YNEV \citep{z15} to test the linear model in overshoot mixing. We use two approaches to calculate the stellar evolutionary models to compare with each other. The first approach is to implement the full nonlocal turbulent convection model (e.g., Eqs.(\ref{TCM_kr}-\ref{TCM_TT}) and the overshoot mixing model Eq.(\ref{z13ovm}) in stellar evolution (see \citet{z15}, Section 3.2), denoted as 'full TCM' approach. The MLT theory is replaced by the turbulent convection model. The convective heat flux determining the temperature gradient and the dissipation rate $\varepsilon$ determining the diffusion coefficient of mixing in overshoot region are calculated by using the turbulent convection model.
The second approach is standard stellar model (use MLT to determine the convective heat flux) with an extra diffusion overshoot mixing (Eq.(3)) with the dissipation rate $\varepsilon$ determined by using the linear model, denoted as 'linear model' approach. .

In the adopted stellar evolution code, the convection zone is artificially fully mixed at first and then we solve the overshoot mixing. This may lead a problem in using Eq.(\ref{z13ovmn2}). Because the artificial mixing may lead to a discontinuity at the convective boundary, $N_{turb}^2$ may not exist at the boundary in that case if using Eq.(\ref{z13ovmn2}) to calculate. In order to avoid this problem resulting from the artificial mixing, we use this formula to calculate $N_{turb}^2$:
\begin{eqnarray} \label{ntb2-used}
{N_{turb}}^2 =  - \frac{{\delta g}}{{{H_P}}}[\nabla  - {\nabla _{ad}} - \\ \nonumber
\psi {C_1}{C_A}\sum\limits_{i = 1}^I {{{(\frac{{\partial \ln T}}{{\partial {X_i}}})}_{P,\rho ,X - \{ {X_i}\} }}\frac{{d{X_i}}}{{d\ln P}}} ]
\end{eqnarray}%
where the parameter $\psi$ is defined as
\begin{eqnarray} \label{ntb2-psi}
\psi  = {\rm{Min}}[1,{\rm{Max}}(0,\frac{{{\nabla _R} - {\nabla _{ad}} - d}}{d})]
\end{eqnarray}%
and $d$ is a small value (we set $d=0.002$) to determine the depth of the swap region between the convective boundary and the location $\psi=1$. Using that formula to calculate $N_{turb}^2$ is reasonable, because the swap region is small and it should be efficiently mixed due to the high diffusion coefficient.
In solving the diffusion equation of mixing:
\begin{eqnarray} \label{difov}
\frac{{\partial {X_i}}}{{\partial t}} = \frac{\partial }{{\partial m}}[{(\frac{{dm}}{{dr}})^2}D\frac{{\partial {X_i}}}{{\partial m}}],
\end{eqnarray}%
the diffusion coefficient $D$ is calculated by using Eq.(\ref{z13ovm}) in overshoot region and its upper limit is set to be $D=10^{10}$ (enough for ensure the full mixing), the boundary conditions are zero flux conditions at the stellar center and the stellar surface.

The OPAL equation of state tables EOS2005 \citep{EOS2005} are used to calculate the thermodynamic functions. The Rosseland mean opacities in high and low temperature region are interpolated from the OPAL tables \citep{OPAL} and the low-temperature tables \citep{F05}, respectively. The rates of all nuclear reactions are based on \citet{Nucl} and \citet{CF88} and enhanced by a factor due to weak electron screening \citep{sal54}. The composition in heavy elements are set as the GN93 \citep{GN93} or AGSS09 \citep{AGSS09} solar composition. Except the overshoot mixing, non-standard physical processes (settling, mass-loss, rotation and etc.) are not taken into account.

In our tests, the basic values of parameters are as follows. The convection parameters are $\alpha_{MLT}=1.75$ for MLT and $\alpha=0.8$ for the TCM. They are the typical value for solar calibrations. Solar calibrations show that $\alpha_{MLT}=(2.1 \sim 2.2)\alpha$. Other TCM parameters are: $C_{t1}=C_{e1}=0$, $C_t=7.5$, $C_e=0.2$, $C_s=0.08$ and $C_k=2.5$. TCM parameters $C_t$, $C_e$, $C_s$ and $C_k$ are based on the reproduction of the temperature gradient below the solar convection zone to the helioseismic required profile \citep{zl12a}. The parameters in the overshoot mixing model are $C_{OV}=10^{-3}$ based on calibrations on some observations \citep{z13,mz14} and $C_1=0.72$ \citep{can98,can11}. In order to test the effects of different values of parameters, the parameters are varied in large ranges around their basic values.

\subsection{validating the linear model}

We calculate stellar evolutionary models with the mass range $1.5M_{\odot}\sim10M_{\odot}$ from ZAMS to the AGB phase (or RGB phase for the $1.5M_{\odot}$ star). We then compare the model properties (e.g., turbulent kinetic energy, diffusion coefficient, abundance in stellar interior and the stellar evolution tracks) between two approaches of calculating the stellar evolutionary models: the 'full TCM' and the 'linear model' approaches. Parameters of the TCM and the overshoot model are set as their basic values.

\begin{figure}
\includegraphics[scale=1.0]{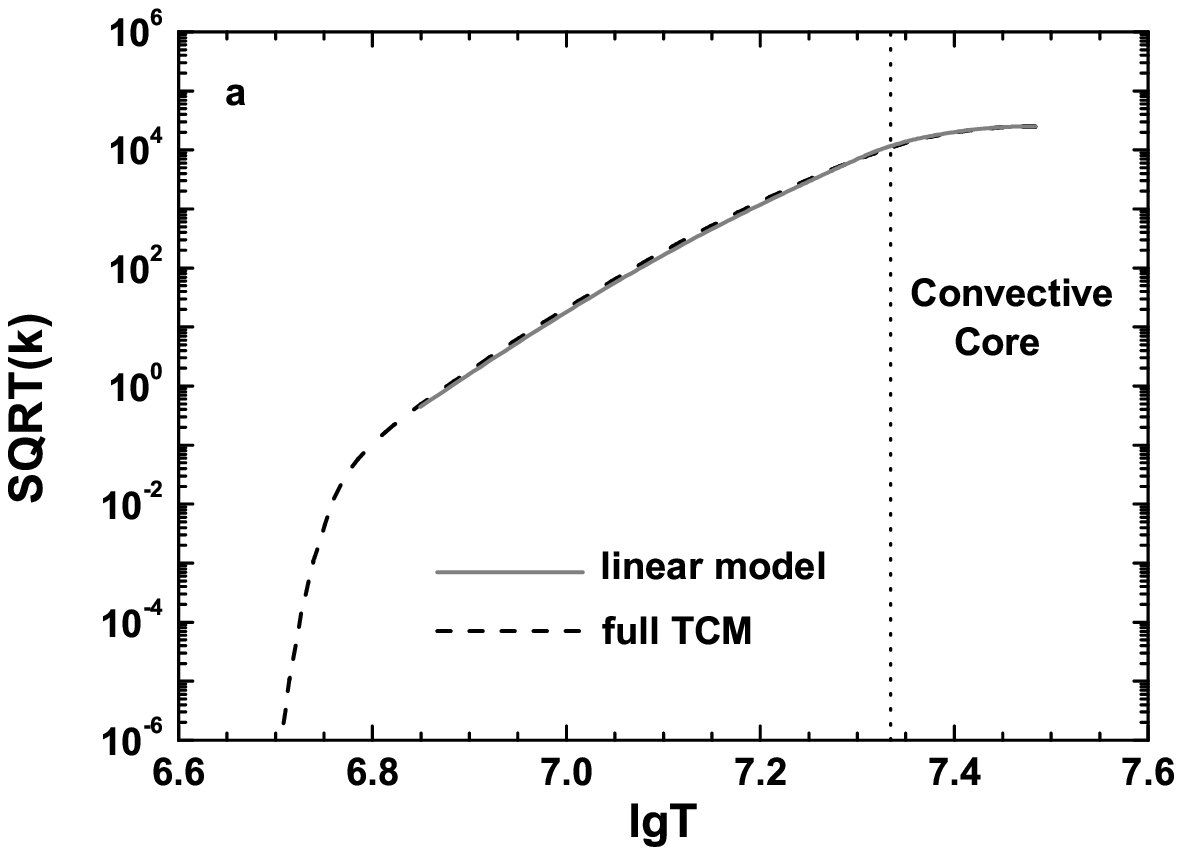}
\includegraphics[scale=1.0]{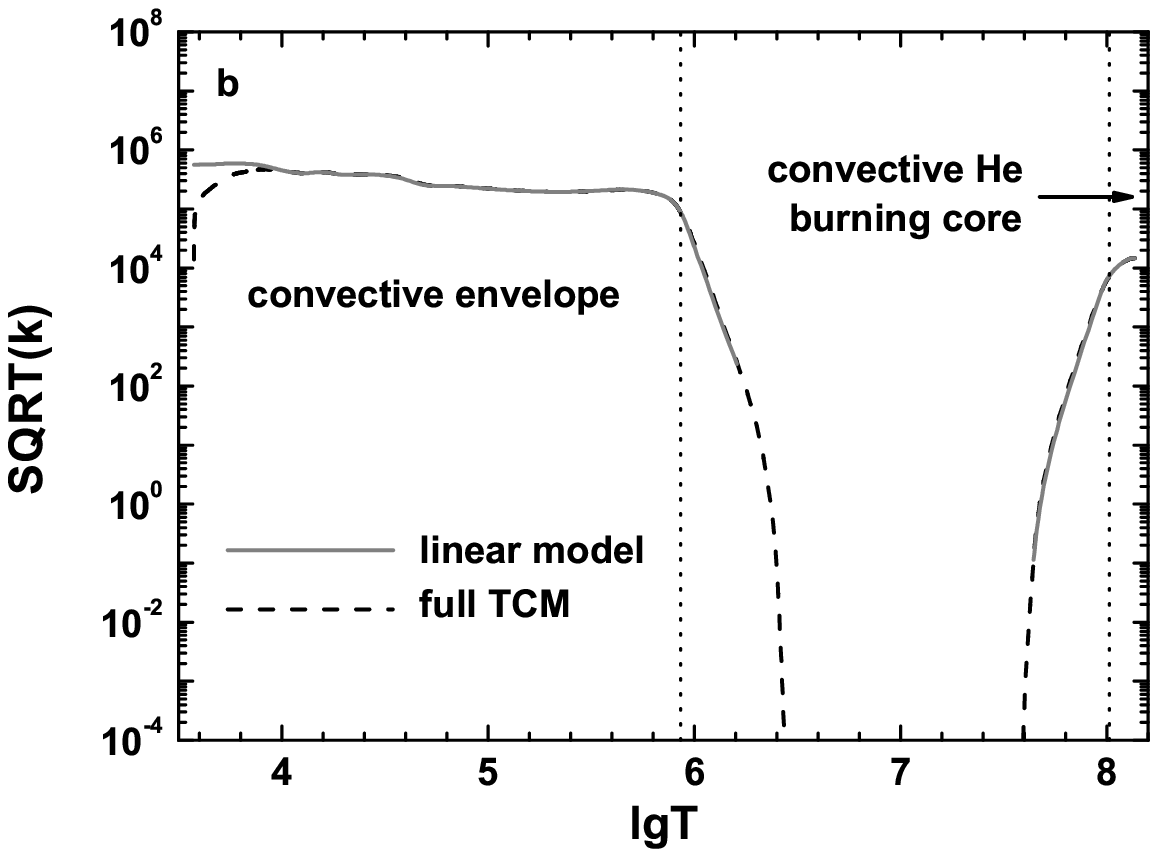}
\caption{Turbulent r.m.s. speed in $cm/s$ for a 7M stellar model with $X=0.7$, $Z=0.02$ and GN93 mixture: a - at the time $X_C=0.4$, b - at the top of RGB. The gray solid lines are calculated by the using linear model of turbulent kinetic energy, and the black dashed lines are calculated by the full nonlocal turbulent convection model. The dotted-lines indicate the convective boundaries. } \label{turbk7mGN93}
\end{figure}

\begin{figure}
\includegraphics[scale=1.0]{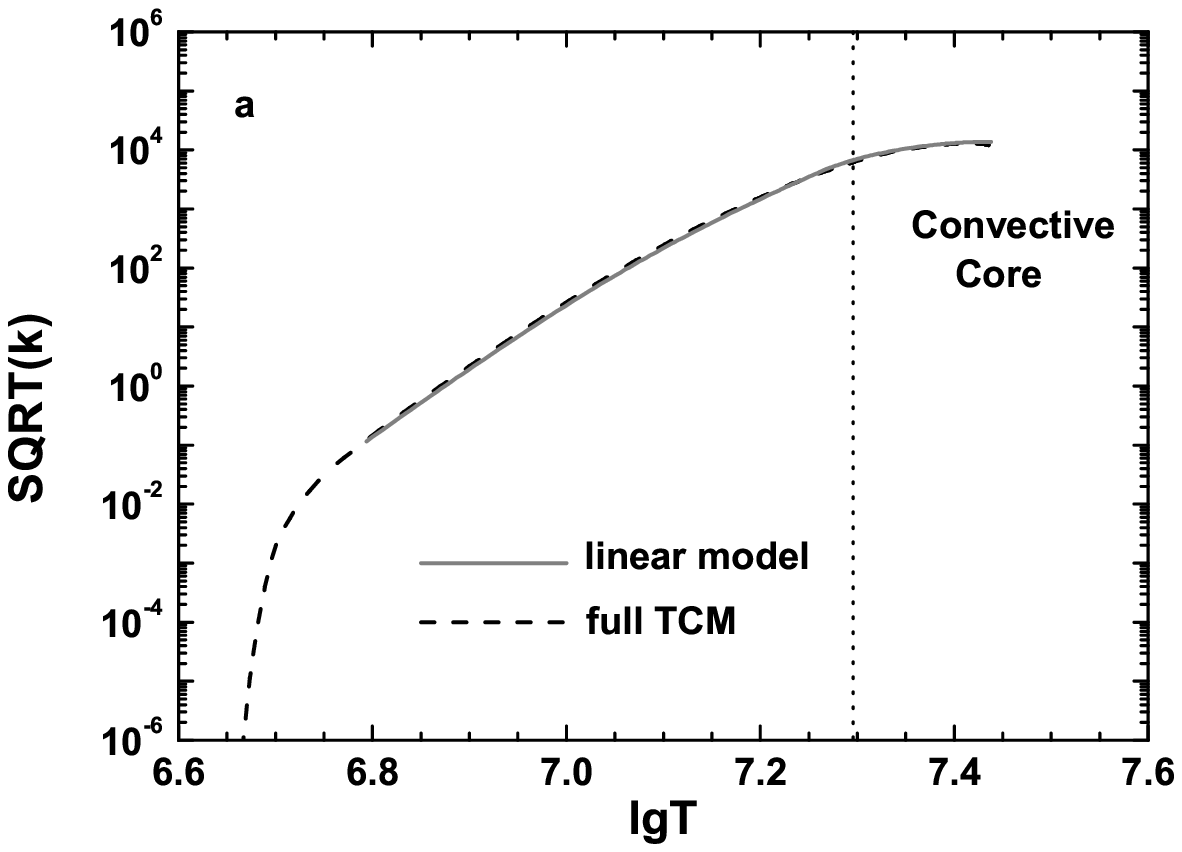}
\includegraphics[scale=1.0]{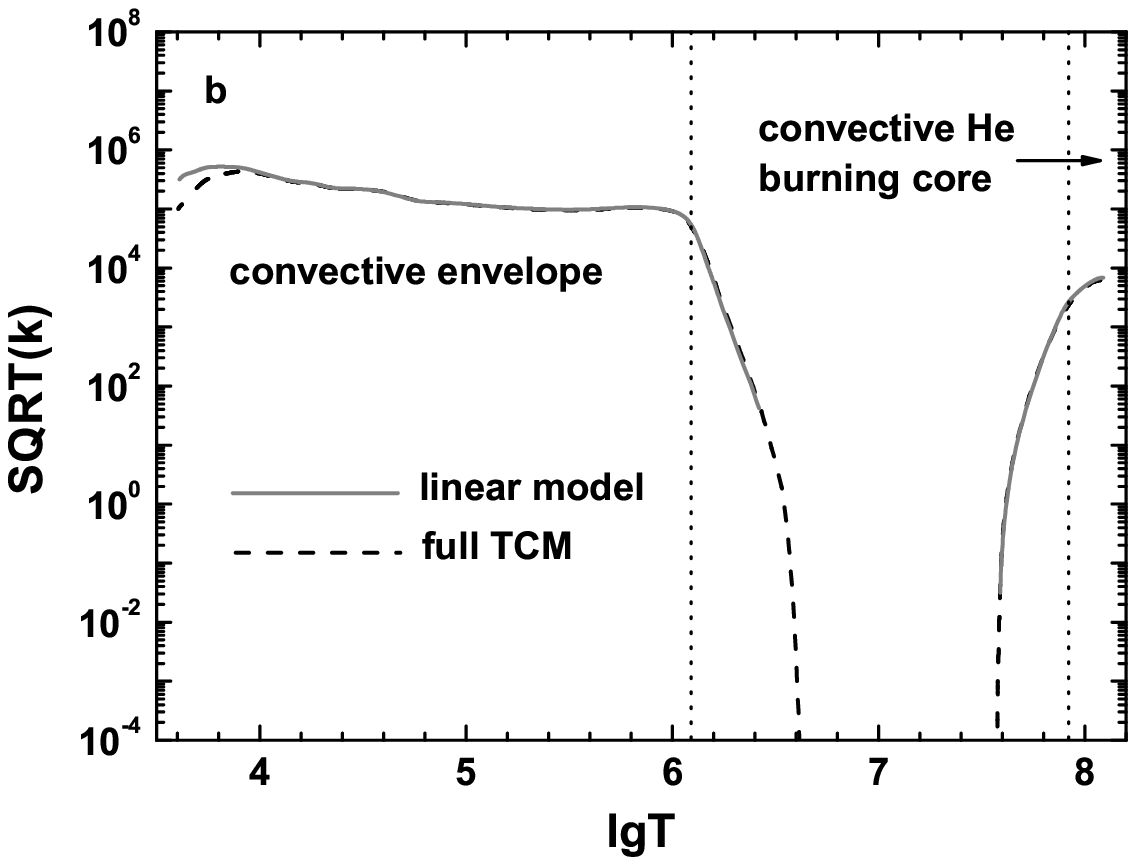}
\caption{Similar to Fig.1, but for a 4M stellar model with $X=0.715$, $Z=0.014$ and AGSS09 mixture: a - at the time $X_C=0.4$, b - at the top of RGB. } \label{turbk4mAGSS09}
\end{figure}

Turbulent kinetic energy $k$ is a direct indicator to check whether the linear model is a good approximation of the full TCM. Figures \ref{turbk7mGN93} and \ref{turbk4mAGSS09} show the profiles of turbulent r.m.s. speed $\sqrt{k}$ obtained by using linear model and full TCM for $4M_\odot$ and $7M_\odot$ stellar models in different composition and state. The stellar models in Figs. \ref{turbk7mGN93}a and \ref{turbk4mAGSS09}a are in main sequence stage with the hydrogen abundance in the center $X_C=0.4$. The turbulent r.m.s. speed profiles obtained by using linear model and full TCM are almost identical, validating the linear model is a reasonable simplification of the full TCM. In the linear model, we cut-off $k$ at about $lgT=6.85$ because the P\'{e}clet number is smaller than unit in the overshoot layer with $lgT<6.85$. The TCM shows as the dashed line that $k$ quickly decreases to zero in the overshoot region with $Pe \ll 1$. The stellar models in Figs. \ref{turbk7mGN93}b and \ref{turbk4mAGSS09}b are at the top of red giant branch where the center helium burning just begins. There is a thick convective envelope and a convective helium burning core in each stellar model. The linear model well reproduces the result of full TCM except the surface layer. The difference is due to $Pe \ll 1$ in the surface layer so that the assumption of linear model does not stand. However, as mentioned in Section 4, the difference exists only in the low P\'{e}clet number layer extending by several $l$ and does not affect the mixing diffusion coefficient in the overshoot regions.

\begin{figure}
\includegraphics[scale=1.0]{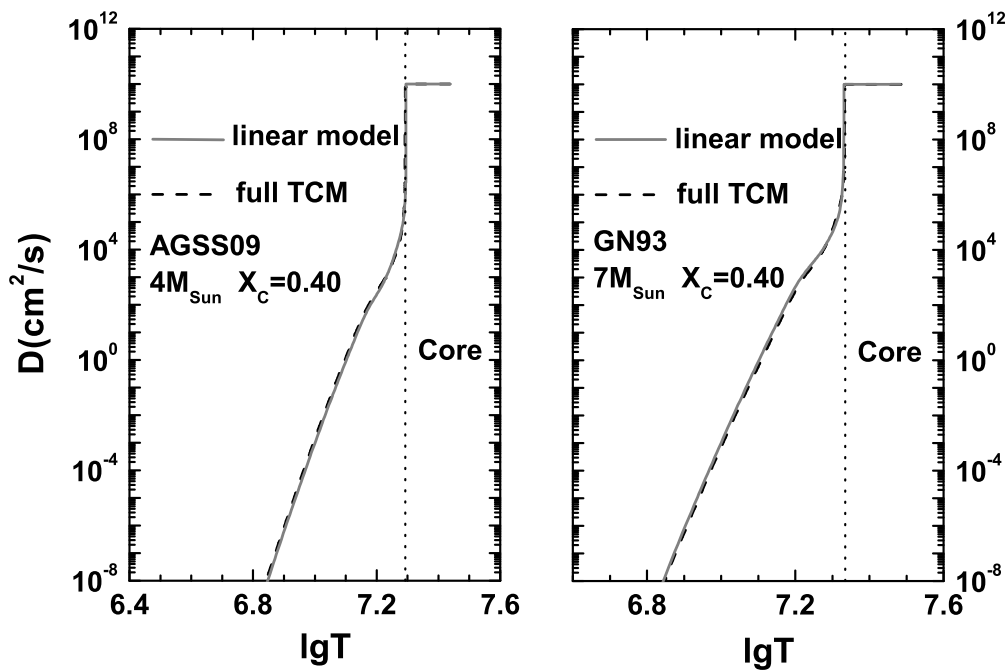}
\caption{Diffusion coefficient in convective core overshoot regions for two 4$M_{\odot}$ stellar models with $X=0.715$, $Z=0.014$ and AGSS09 mixture and two 7$M_{\odot}$ stellar models with $X=0.7$, $Z=0.02$ and GN93 mixture at the time $X_C\approx0.4$. The gray solid lines are for the stellar models calculated by the using linear model of turbulent kinetic energy and the overshoot mixing model, and the black dashed lines are the stellar models calculated by the full nonlocal turbulent convection model and the overshoot mixing model. The dotted-lines indicate the convective boundaries. } \label{difc}
\end{figure}

\begin{figure}
\includegraphics[scale=1.0]{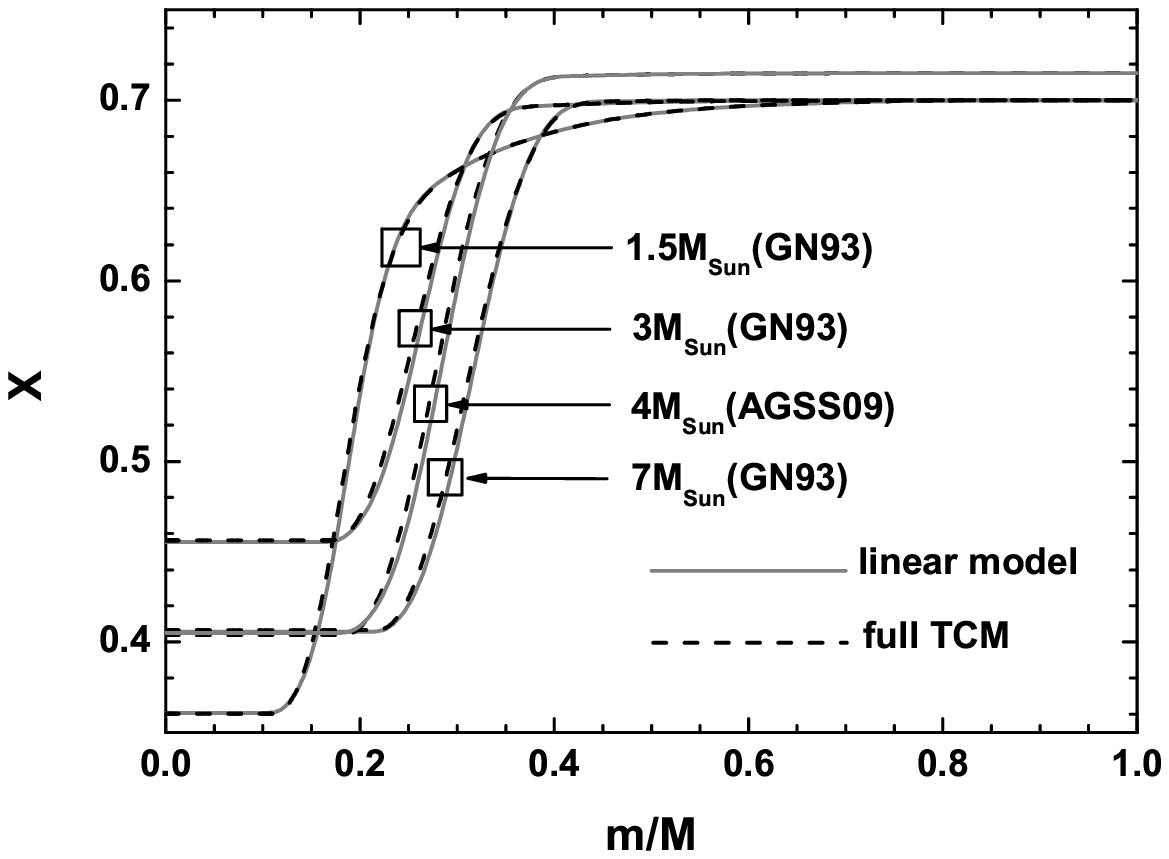}
\caption{Hydrogen abundance in stellar interior for stellar models with $X=0.7$, $Z=0.02$ and GN93 mixture: two 1.5$M_{\odot}$ stellar models with $X_C\approx0.36$, two 3$M_{\odot}$ stellar models with $X_C\approx0.46$ and two 7$M_{\odot}$ stellar models with $X_C\approx0.4$. Two 4$M_{\odot}$ stellar models with $X=0.715$, $Z=0.014$ and AGSS09 mixture at the time $X_C\approx0.4$ are also shown. The gray solid lines are for the stellar models calculated by the using linear model of turbulent kinetic energy and the overshoot mixing model, and the black dashed lines are the stellar models calculated by the full nonlocal turbulent convection model and the overshoot mixing model. } \label{chem}
\end{figure}

The profiles of diffusion coefficient of convective mixing for main sequence stellar models with different mass are shown in Fig. \ref{difc}. The result of linear model is almost identical to the result of full TCM. We have not calculated the diffusion coefficient in the convective core but set to be $10^{10}$ which is large enough to ensure complete mixing. According to the convective mixing model \citep{z13}, the diffusion coefficient in the convective core is $ \sim l\sqrt{k}$ which is much larger than $10^{10}$ and leads to a fully mixed core. The convective mixing model shows that the form of diffusion coefficient transfers from $D\sim l\sqrt{k}$ in the convective core to $D \approx C_{OV} \varepsilon / N_{turb}^2$ in the overshoot region \citep{z13}. This is shown in the figure that the diffusion coefficient quickly decreases near the convective boundary and exponentially decreases in the overshoot region. The latter is because $\varepsilon$ exponentially decreases and $N_{turb}^2$ changers much slower than $\varepsilon$ in most part of the overshoot region. The diffusion coefficient in the overshoot region is not high enough to ensure complete mixing. This leads to smooth profile of abundance in stellar interior, as shown in Fig. \ref{chem}. The classical overshoot mixing model, which extends the fully mixing region from convective boundary by a distance, can not obtain such smooth profile of abundance. Figure \ref{chem} shows hydrogen abundance profile in stellar interior for different mass in main sequence. The nearly identical profiles validate that the linear model can be used to substitute the full TCM in modeling overshoot mixing.

\begin{figure*}
\includegraphics[scale=1.2]{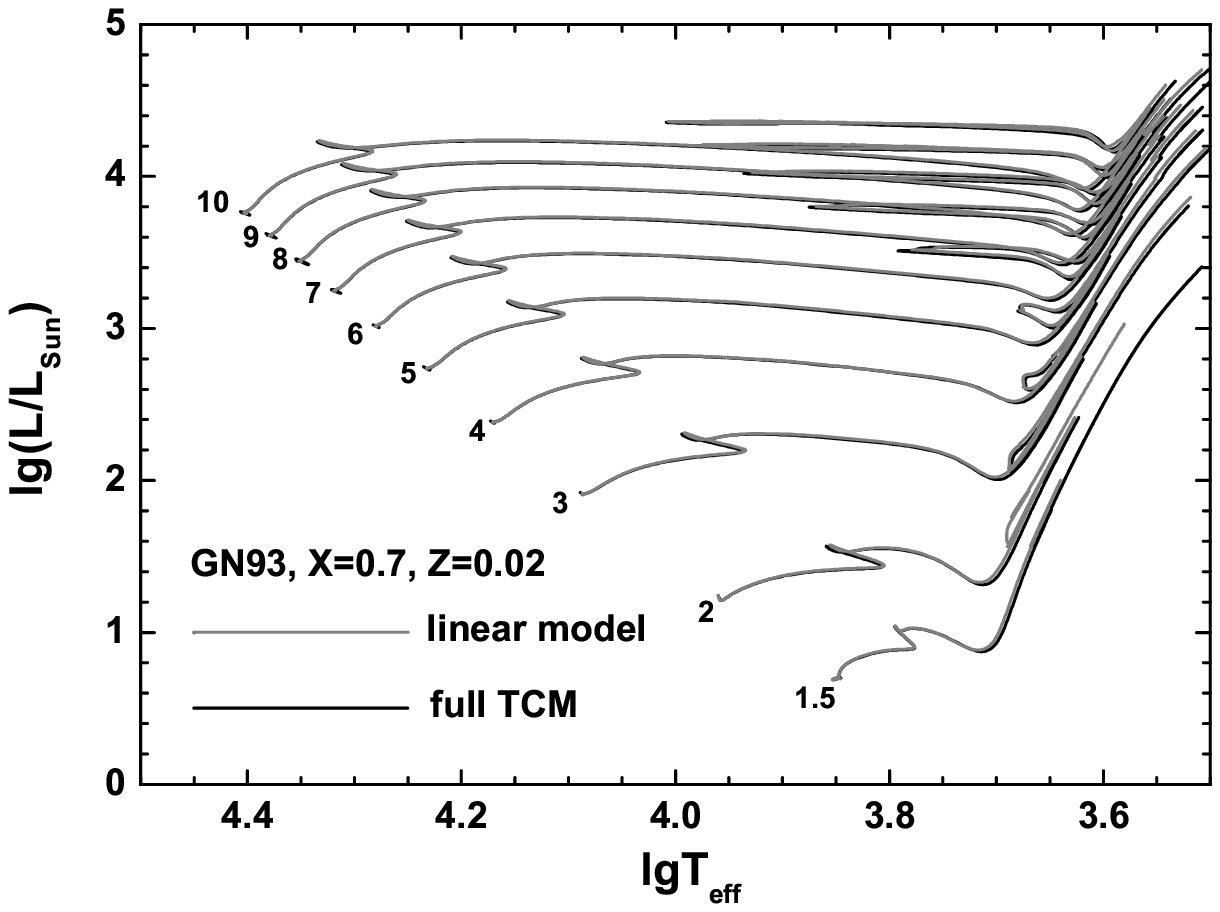}
\caption{Stellar evolutionary tracks in HR diagram for stellar models with $X=0.7$, $Z=0.02$ and GN93 mixture. The gray lines are calculated by the using linear model of turbulent kinetic energy and the overshoot mixing model, and the black lines are calculated by the full nonlocal turbulent convection model and the overshoot mixing model. } \label{tracksGN93}
\end{figure*}

\begin{figure*}
\includegraphics[scale=1.2]{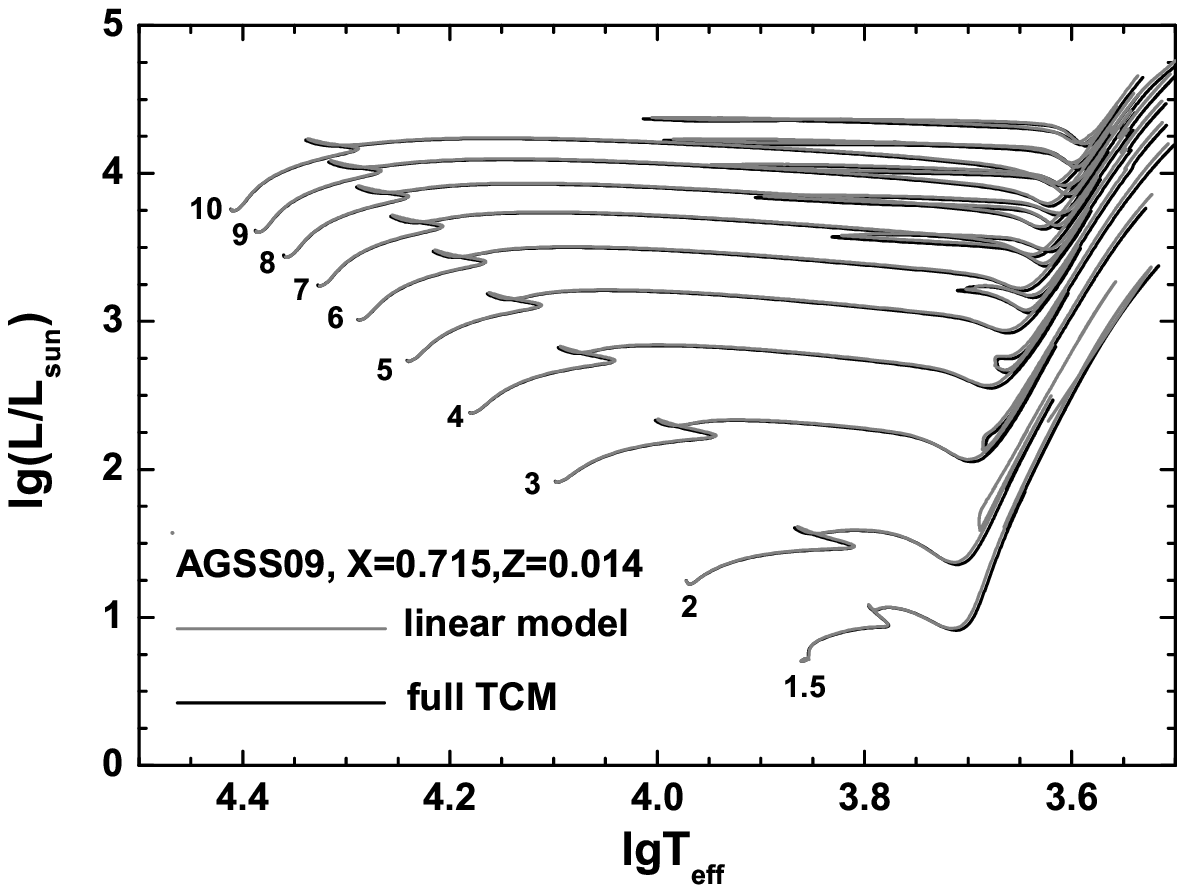}
\caption{Similar to Fig.5, but for stellar models with $X=0.715$, $Z=0.014$ and AGSS09 mixture. } \label{tracksAGSS09}
\end{figure*}

The stellar evolutionary tracks with linear model and full TCM for different mass are shown in Fig. \ref{tracksGN93} and Fig. \ref{tracksAGSS09}. The track with linear model are almost identical to the track with full TCM in the all stage for different stellar mass. This validates that the linear model results in the same strength of overshoot mixing comparing with the full TCM. In the red giant branch, there is a small difference on temperature (about $50K$) between stellar models with linear model and full TCM. This is caused by the difference of turbulent heat transport efficiency between TCM and MLT (adopted in the stellar models with linear model) in super-adiabatic convection zone near the surface. In the solar case, in order to generate similar turbulent heat transport efficiency in super-adiabatic convection zone, we requires $\alpha_{MLT}=(2.1 \sim 2.2)\alpha$ as mentioned above. However, this ratio may change a little for different stellar mass or different evolutionary stage. This effect is ignorable for intermediate mass main sequence stars because the thin convective envelope is dominated by the radiative heat transport and the convective core is almost adiabatic stratified, regardless of which convection theory (MLT or TCM) is adopted.

\subsection{Effects of the parameters}

\begin{figure*}
\includegraphics[scale=1.2]{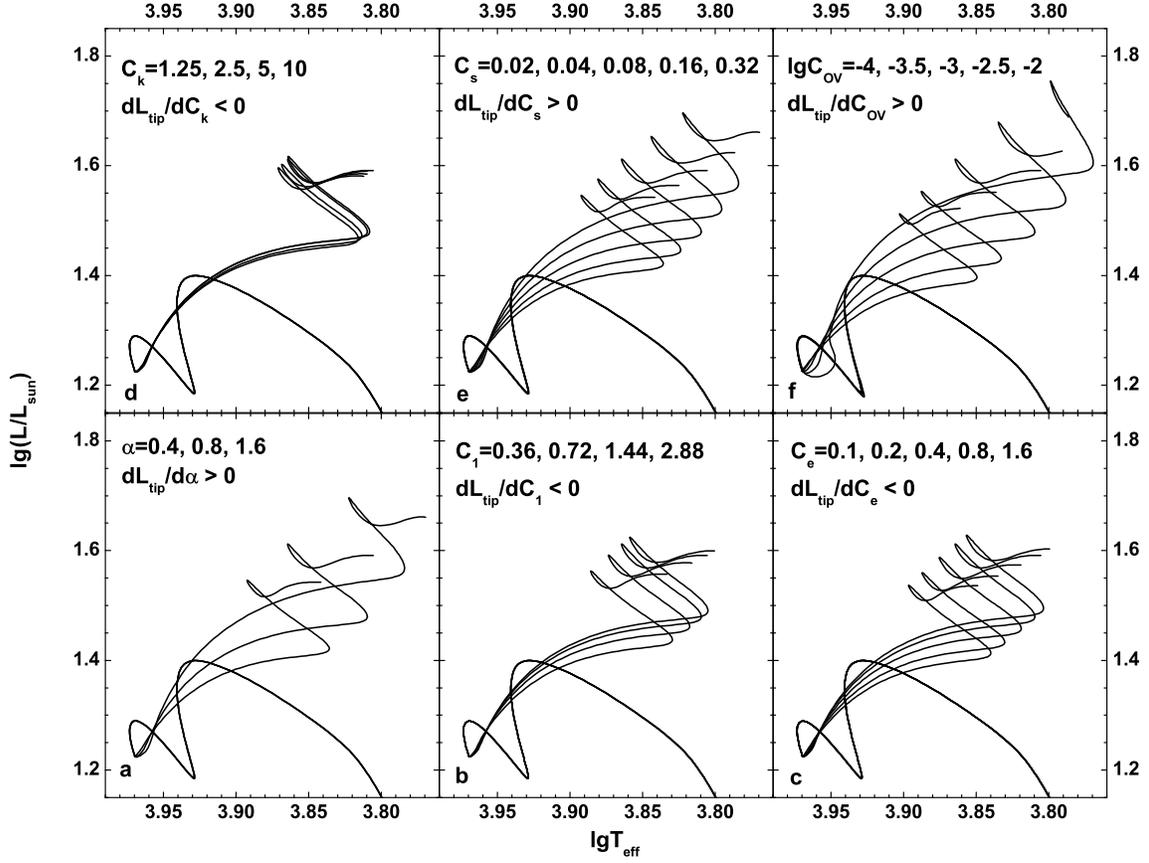}
\caption{Evolutionary tracks for $2M_{\odot}$ stellar models with $X=0.715$, $Z=0.014$ and AGSS09 mixture. The parameters $\alpha$, $C_e$, $C_k$, $C_s$, $C_1$ and $C_{OV}$ are varied around their basic values. $L_{tip}$ is the luminosity at the tip. For a parameter $C$, the sign of $dL_{tip}/dC$ indicates the value of the parameter for each evolutionary track. } \label{tracksPara}
\end{figure*}

Parameters involved in the linear model of nonlocal turbulent convection model are $C_s$, $C_e$, $C_k$ and $\alpha$, and parameters involved in the overshoot mixing model are $C_{OV}$ and $C_1$. We have tested the effects of those parameters on $2M_{\odot}$ ($X=0.715$, $Z=0.014$, AGSS09 composition) stellar models.
Figure \ref{tracksPara} shows the evolutionary tracks in main sequence with different value of parameters (for the parameters which are not specific, they are set as their basic values: $C_e=0.2$, $C_s=0.08$, $C_k=2.5$, $\alpha=0.8$, $C_1=0.72$ and $C_{OV}=10^{-3}$). It is found that the overshoot mixing is enhanced when $C_s$, $\alpha$ or $C_{OV}$ becomes larger or $C_1$ or $C_e$ becomes smaller. Figure \ref{tracksPara}d shows that tracks are insensitive to the parameter $C_k$. In Fig.\ref{tracksPara}a and Fig.\ref{tracksPara}e, the track for $C_s=0.08$ is same one to $\alpha=0.8$, the track for $C_s=0.02$ is almost identical to the track for $\alpha=0.4$ and the track for $C_s=0.32$ is almost identical to the track for $\alpha=1.6$. Those seem to imply that changing $C_s$ by a factor $a$ is almost equivalent to change $\alpha$ by the factor $\sqrt{a}$.

Those properties can be explained as follows. Since the linear model and the overshoot mixing model affect the stellar evolutionary tracks via the diffusion coefficient, let's analyze the dependence of diffusion coefficient on parameters.
As it is shown in Eq.(\ref{z13ovm}), diffusion coefficient is in proportion to the turbulent dissipation rate which is determined by the linear model. By using Eq.(\ref{kOV}), Eq.(\ref{ksita}) and Eq.(\ref{kcb}), we find:
\begin{eqnarray} \label{epsi-analysis}
\ln \varepsilon  &=& \ln \frac{{{k^{\frac{3}{2}}}}}{l} \approx \ln \{ \frac{{[\delta g{D_R}({\nabla _R} - {\nabla _{ad}})]_B}}{{e{H_P}}}{(\frac{P}{{{P_C}}})^{\frac{3}{2}\theta }}\}  \\ \nonumber
&\approx& {\rm{const}}{\rm{.}} + f({\alpha ^2}{C_s}) + \frac{3}{2}\theta({\alpha ^2}{C_s}, C_e, C_k) \ln (\frac{P}{{{P_C}}}) \\ \nonumber
&=& {\rm{const}}{\rm{.}} + f({\alpha ^2}{C_s}) + \frac{3}{2}\sqrt {\frac{{1 + 2{C_e}{\omega _O}}}{{3{\alpha ^2}{C_s}{\omega _O}}}} \ln P,
\end{eqnarray}%
where the function of ${\alpha ^2}{C_s}$ term $f({\alpha ^2}{C_s})$ represents the location of the point B depending on ${\alpha ^2}{C_s}$ (see Eq.(\ref{rb})).
For the case of core overshoot, the positive value of $\theta$ is adopted and there is $f'>0$ in general case because local kinetic energy in the convective core decreases toward the Schwarzchild boundary.
It should be noticed that two parameters ${\alpha }$ and ${C_s}$ can be combined to one. This is the reason for changing $C_s$ by a factor $a$ being almost equivalent to changing $\alpha$ by the factor $\sqrt{a}$. It is not difficult to find that $d\theta/d({\alpha ^2}{C_s})<0$ and $d\theta/dC_e>0$. This means that the exponential index of the diffusion coefficient becomes smaller when $({\alpha ^2}{C_s})$ becomes larger or $C_e$ becomes smaller. A smaller index of diffusion coefficient leads to higher efficiency for the mixing. $C_k$ does affect $\omega _O$ only. For the testing range of $C_k$, $\omega _O$ changes a little so that the tracks are insensitive to $C_k$. When $C_1$ becomes larger, the weight of abundance gradient in $N^2_{turb}$ is larger so that $N^2_{turb}$ increases and the diffusion coefficient decreases. The diffusion coefficient is in proportion to the parameter $C_{OV}$ so that the overshoot mixing is enhanced as $C_{OV}$ increases.

\subsection{Effects of the modification of temperature gradient in overshoot region}

The convective heat flux could modify the temperature gradient in the overshoot region. This affects the value of ${N_{turb}}^2$ and thus affects the diffusion coefficient of overshoot mixing. In the previous calculations of stellar models based on 'linear model', this effect was not taken into account since the temperature gradient is calculated in the traditional way (MLT in convection zones and the radiative temperature gradient is adopted outside convection zones). In this subsection, we investigate the effects of taking into account the modification of temperature gradient in the overshoot region.

We have implemented the modification of temperature gradient in the overshoot region as follows: step 1, solving the linear model to work out / update the convective heat flux in overshoot region (e.g., Eq(\ref{u't'OV})) before every iteration in solving the stellar structure equations; step 2, solving the stellar structure equations by one iteration with the updated convective heat flux in overshoot region. The iterations stop when the required accuracy is achieved.

It is not difficult to understand that the negative convective heat flux in the overshoot region should enlarge the temperature gradient and make it closer to the adiabatic temperature gradient, thus reduce ${N_{turb}}^2$ and enlarge the diffusion coefficient of mixing. It can be found in Eq.(\ref{lad}) that enlarging parameters $\alpha^2C_s$ or $C_e$ enhances the modification on temperature gradient. At here, we take a large value of $C_e=0.8$ for example. Other parameters are set as their basic values. We have tried $C_e=1.6$ but we can not get converged stellar models.

\begin{figure}
\includegraphics[scale=1.0]{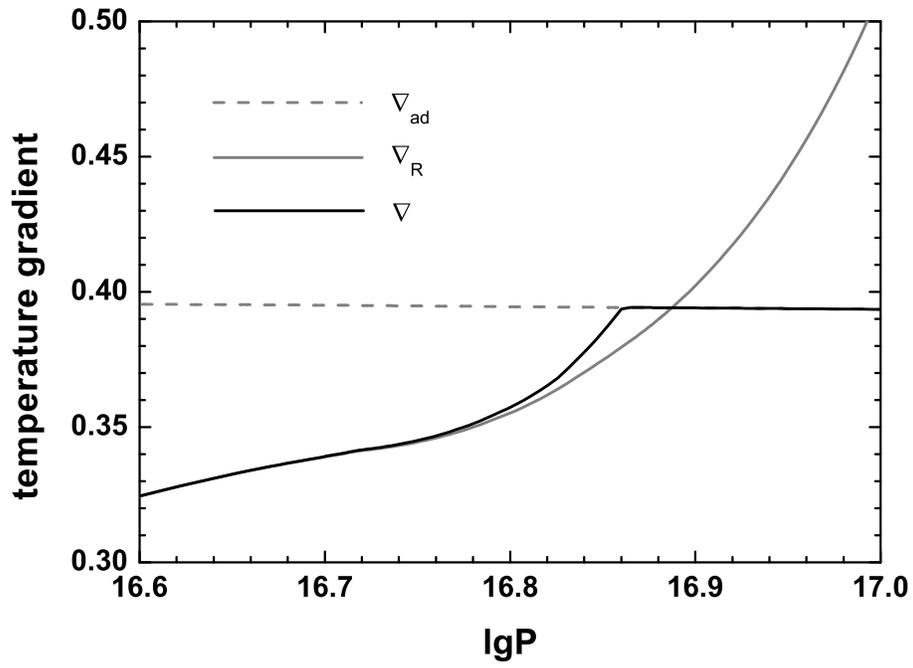}
\caption{ Temperature gradient near the convective core boundary for a 4$M_{\odot}$ stellar model with $X=0.715$, $Z=0.014$ and AGSS09 mixture at $X_C\approx0.32$. The black solid line is temperature gradient $\nabla$, the gray solid line is the radiative temperature gradient $\nabla_R$ and the gray dashed line is the adiabatic temperature gradient $\nabla_{ad}$. } \label{GRT}
\end{figure}

\begin{figure}
\includegraphics[scale=1.0]{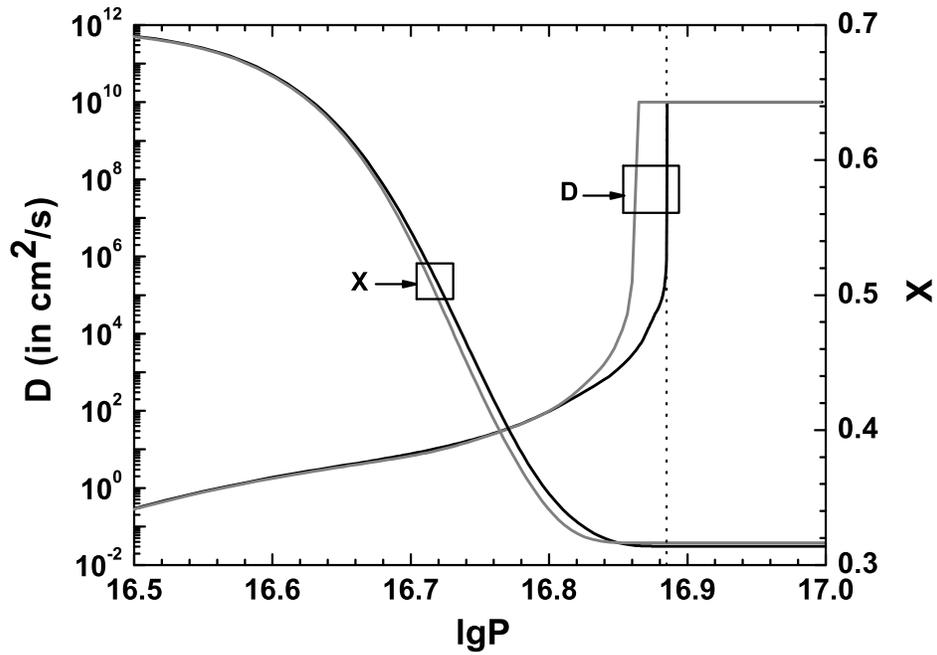}
\caption{ Diffusion coefficient and the hydrogen abundance near the convective core boundary for a 2$M_{\odot}$ stellar model with $X=0.715$, $Z=0.014$ and AGSS09 mixture at $X_C\approx0.32$, with and without the temperature gradient modification. The black lines are for the stellar model without temperature gradient modification and the gray lines are for the stellar model with temperature gradient modification. }  \label{GRT_DX}
\end{figure}

\begin{figure}
\includegraphics[scale=1.0]{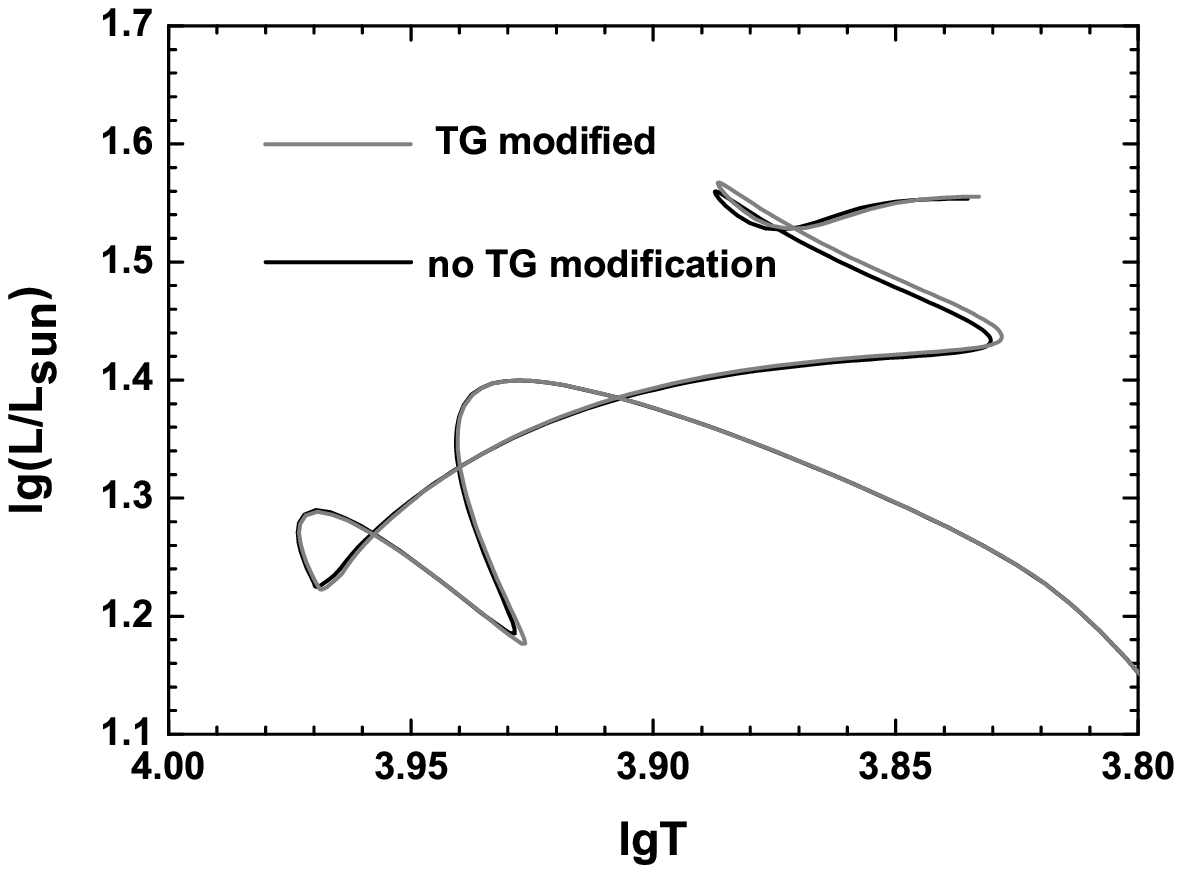}
\caption{ Evolutionary tracks of 2$M_{\odot}$ stellar models with $X=0.715$, $Z=0.014$ and AGSS09 mixture with and without the temperature gradient modification. The black line are for the stellar models without temperature gradient (TG) modification and the gray line are for the stellar models with temperature gradient modification. } \label{GRT_EVOL}
\end{figure}

Based on the linear model, 2$M_{\odot}$ stellar models with $X=0.715$, $Z=0.014$ and AGSS09 mixture have been calculated, with or without the modification of temperature gradient in the overshoot region. Figure \ref{GRT} shows the temperature gradient near the convective core boundary for the stellar model with the modification at the moment $X_C\approx0.32$. Temperature gradient modification is significant in a range $\sim0.1H_P$ outside the convective core. Figure \ref{GRT_DX} shows the diffusion coefficient and hydrogen abundance profile of that stellar model, as well as the stellar model without the modification of temperature gradient. It can be found that diffusion coefficient near the convective boundary has been enlarged due to the modified temperature gradient. However, the difference between hydrogen abundance profiles of the two models is not significant. This is because the diffusion coefficient near the convective boundary is very high so that the abundance is very close to the abundance in the core as shown in the figure. In this case, enlarging the diffusion coefficient has little effects. The inhomogenous region is in low diffusion coefficient region, in which the convective heat flux is ignorable so that the diffusion coefficient is not affected. Therefore the hydrogen abundance profiles of the two models are very close to each other. Figure \ref{GRT_EVOL} shows the evolutionary tracks for the stellar models with and without the modification of temperature gradient in the overshoot region. It is also shown that, although the modification can enhance the overshoot mixing, the effect is not significant. The comparison is for a large value of $C_e=0.8$. If the basic value is adopted, the effect should be less important.

\section{Summary}

Asteroseismic studies do not support the classical 'ballistic' overshoot model and implies that the convective overshoot is a weak mixing process. \citeauthor{z13}'s \citeyearpar{z13} overshoot mixing model describes the overshoot as a weak mixing process and the diffusion coefficient shows consistence with the overshoot entropy mixing. However, we need to know the dissipation rate of turbulent kinetic energy in overshoot region before applying that mixing model. A practicable option to work out the dissipation rate is to solve nonlocal turbulent convection models (TCMs), but this is difficult because of some numerical problems and the time costs is hard to bear.

In this paper, we have simplified the full nonlocal turbulent convection model developed by \citet{li07} to a linear model (e.g., Eqs. (\ref{est_kc}), (\ref{u't'linear-a}), (\ref{u't'linear-b}), (\ref{omi-a}) and (\ref{omi-b})) in order to obtain the dissipation rate of turbulent kinetic energy which is required in the overshoot mixing model. The linear model is a single linear diffusion equation for turbulent kinetic energy. It is very easy to be implemented in a stellar evolution code. The time cost of solving the linear model is ignorable to compare with solving the full TCM. And there is no numerical difficulty in solving the linear model. We have tested the linear model in stellar evolution code, and have found that the linear model can well reproduce the turbulent kinetic energy profile of full TCM, as well as the diffusion coefficient, abundance profile and the stellar evolutionary tracks. We have also studied the effects of different values of the model parameters and have found that the effect due to the modification of temperature gradient in the overshoot region is slight.

\acknowledgments

Many thanks to the anonymous referee for careful reading of the manuscript and providing comments which improved the original version. Fruitful discussions with Y. Li are highly appreciated.
This work is co-sponsored by the National Natural Science Foundation of China through grant No. 11303087 and the Chinese Academy of Sciences ("Light of West China" Program and Youth Innovation Promotion Association).

\appendix

\section{Analysis of the root of the quadratic equation of $\omega _O$}

The asymptotical equilibrium value of the anisotropic degree in overshoot region $\omega _O$ satisfies the following equation:
\begin{eqnarray} \label{appendixA-omio}
2{C_e}{\omega _O}^2 - ({C_k} - 1 + 2{C_e}){\omega _O} + \frac{1}{3}({C_k} - 1) = 0
\end{eqnarray}%
For convenient, we define $c=(C_k-1)/(2C_e)$ thus the equation above can be written as:
\begin{eqnarray} \label{appendixA-omio2}
F({\omega _O}) \equiv {\omega _O}^2 - (1 + c){\omega _O} + \frac{c}{3} = 0
\end{eqnarray}%

Since the discriminant is always positive, let $\omega _1$ and $\omega _2$ be the two roots of the quadratic equation and $\omega _1<\omega _2$.
According to Vieta's theorem, there are:
\begin{eqnarray} \label{appendixA-Vieta-multiply}
{\omega _1}{\omega _2} = \frac{c}{3},
\end{eqnarray}%
\begin{eqnarray} \label{appendixA-Vieta-plus}
{\omega _1} + {\omega _2} = 1 + c.
\end{eqnarray}%
The values of function $F$ at 0, $1/3$ and 1 are:
\begin{eqnarray} \label{appendixA-values}
F(0) = \frac{c}{3},F(\frac{1}{3}) =  - \frac{2}{9},F(1) =  - \frac{{2c}}{3}.
\end{eqnarray}%

1. case A: $C_k\leq1$.
In this case, $c\leq0$ so that $\omega _1\leq0\leq\omega _2$. Because $F(1/3)F(1)\leq0$, we find $\omega _2\geq1/3$.
Physically acceptable root should be $0<\omega _O<1/3$ because the work of buoyancy on the radial turbulent kinetic energy is negative in overshoot region, so that there is no acceptable root when $C_k\leq1$.

2. case A: $C_k>1$.
In this case, $c>0$, ${\omega _2} > ({\omega _1} + {\omega _2})/2 = (1 + c)/2 > 1/2$ so that ${\omega _2}$ is not acceptable. Because $F(0)F(1/3)<0$, we find $0<\omega _1<1/3$ which is physically acceptable root.

Finally, $C_k$ must be larger than 1 and the physically acceptable root is the small one.

\end{document}